\documentclass[prl,twocolumn,superscriptaddress,showpacs,floatfix]{revtex4-1}
\usepackage{graphicx}
\usepackage{dcolumn}
\usepackage{bm}
\usepackage{amsmath}
\usepackage{float}
\usepackage{subfigure}

\begin{document}
\title {Construction and Optimization of the Quantum Analog of Carnot Cycles}
\author {Gaoyang Xiao and Jiangbin Gong}
\email[]{phygj@nus.edu.sg}
\affiliation{Department of Physics and Centre for Computational Science and Engineering, National University of Singapore, Singapore 117542}
\date{\today}

\begin{abstract}
The quantum analog of Carnot cycles in few-particle systems consists of two quantum adiabatic steps and two isothermal steps.  This construction is formally justified by use of a minimum work principle.
It is then shown, without relying on any microscopic interpretations of work or heat, that the heat-to-work efficiency of the quantum Carnot cycle thus constructed may be further optimized, provided that two conditions regarding the expectation value of some generalized force operators evaluated at equilibrium states are satisfied. In general the optimized efficiency is system-specific, lower than the Carnot efficiency, and dependent upon both temperatures of the cold and hot reservoirs.  Simple computational examples are used to illustrate our theory. The results should be an important guide towards the design of favorable working conditions of a realistic quantum heat engine.
\end{abstract}
\maketitle

{\it Introduction} -- The big {\it energy} challenge of this century calls for diversified energy research, including a bottom-up approach towards energy efficiency.  Apart from two stimulating implementations of  microscale heat engines \cite{engine1,engine2},  some theoretical aspects as well as
possible realizations of nanoscale heat engines \cite{Bender.00.JPAMG,low1,low3,Abah.12.PRL,Bergenfeldt.14.PRL,Zhang.14.PRL,prl2,Quan.05.PRE,Rezek.06.NJP,Quan.07.PRE,Arnaud.08.PRE, Agarwal.13.PRE,dio,Dario} have been studied.  For purely quantum heat engines at the nanoscale where the working medium may consist of few particles only (e.g., few trapped ions \cite{Abah.12.PRL}), both quantum fluctuations and thermal fluctuations become significant.  General understanding of the design of such energy devices are also of fundamental interest to nanoscale thermodynamics \cite{Jarzynski.97.PRL,Jarzynski.97.PRE,Crooks.99.PRE,hanggireview}.
In particular, as the size of the working medium shrinks to a quantum level, one must reexamine the implications of the second law of thermodynamics for the efficiency of quantum heat engines.  To that end, we construct and look into the quantum analog of Carnot cycles \cite{Carnot.1824.R,Zemansk.1997.H}.

The construction of the quantum analog of a Carnot cycle is not as straightforward as it sounds.
Consider first the two quasi-static isothermal steps during which the working medium is in thermal equilibrium with a reservoir. Regardless of the size of the quantum medium, its thermodynamic properties can therefore be well defined in the standard sense.  As such isothermal steps can be directly carried over to the quantum case.
However, translating the two adiabatic steps of a Carnot cycle into a quantum analog is by no means obvious.
One intuition \cite{Bender.00.JPAMG,prl2,Quan.05.PRE} is to replace quasi-static adiabatic steps in thermodynamics (without heat exchange) by quantum adiabatic processes (as defined in the celebrated quantum adiabatic theorem \cite{adiabatic}). The starting point of this work is to formally justify such an intuitive construction by revealing a fundamental reason related to energy efficiency.

Below we simply call the quantum anolog of a classical Carnot cycle (two isothermal steps and two quantum adiabatic processes) as a quantum Carnot cycle.  It is yet fundamentally different from a conventional Carnot cycle. During the two adiabatic steps,
the working medium implementing the quantum Carnot cycle is generically not at equilibrium conditions, except for the case in which all the energy levels of the working medium are scaled by a common factor as a system parameter varies (to be elaborated below).  Thus, it becomes important to lay out general designing principles concerning  how the efficiency of a quantum Carnot cycle can be optimized, preferably using standard definitions of work and heat. The explicit optimization conditions are presented below.
Our theory also shows that in general the optimized efficiency attained by a quantum Carnot cycle is (i) lower than the standard Carnot efficiency, (ii) not a simple function of ${T_c}/{T_h}$ but a function of both $T_c$ and $T_h$, the temperatures of cold and hot reservoirs, and (iii) depends on the detailed spectrum of the working medium.  These features will guide us in the design of favorable working conditions of a realistic quantum heat engine.
Simple computational examples are used to illustrate our theory.  Throughout this work we do not
use recent microscopic interpretations or definitions of work or heat proposed for quantum systems, such as those introduced in Refs.~\cite{Quan.05.PRE,Quan.07.PRE,prl1,prl2}.   Instead, we only assume that heat exchange is zero if the working medium is thermally isolated and work is zero if the system parameters of the working medium are fixed.

\begin{figure}
\centering
\includegraphics[width=9.4cm]{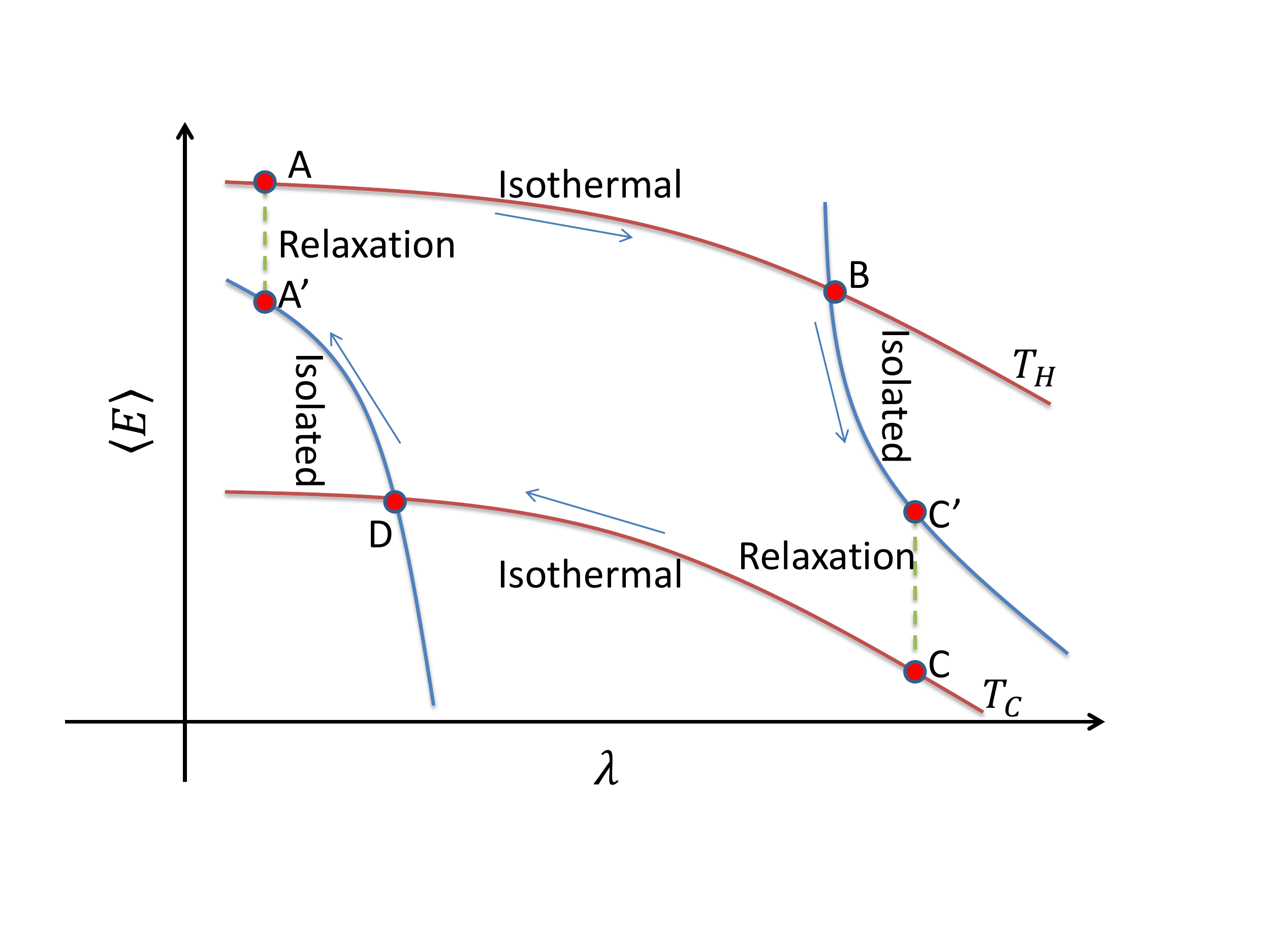}\\
\caption{(Color online) A quantum heat enegine cycle consisting of two isothermal steps and two thermally isolated and hence unitary steps. $A$, $B$, $C$, and $D$ represent four equilibrium states, $C'$ and $A'$ represent two
non-equilibrium states at the end of two unitary steps, approaching respectively to equilibrium states $C$ and $A$ after a relaxation step initiated by contact with the reservoirs.  As shown in the text, for a quantum analog of the Carnot cycle, the two thermally isolated steps should be two quantum adiabatic processes.
}\label{Fig1}
\end{figure}

{\it  Efficiency of quantum heat engine cycles and the second law} --
We start with general considerations of a quantum heat engine cycle consisting of two isothermal steps and two thermally isolated processes.  Figure \ref{Fig1} schematically depicts such a cycle.  There $A\rightarrow B$ and $C\rightarrow D$ represent two isothermal processes during which the quantum medium is always at equilibrium with a reservoir, $\lambda$ is assumed to be the only system parameter tunable in a cycle opeation, $\langle E\rangle$ is the mean energy of the system,
$B\rightarrow C^\prime$ and $D\rightarrow A^\prime$ represent two thermally isolated and hence unitary processes.  The symbols $A^\prime$ and $C^\prime$ are to indicate that right after a unitary process, the quantum medium is in general not at thermal equilibrium.  States of $A^\prime$ and $C^\prime$ will reach thermal equilibrium states $A$ and $C$ after relaxation with a reservoir under fixed values of $\lambda$.

When the system at the nonequilibrium state $C^\prime$ starts heat exchange with the cold reservoir under fixed $\lambda=\lambda_C$, no work is done.  Hence $\langle E_{C^\prime}\rangle - \langle E_C\rangle$ is simply the heat dumped (which could be negative).  This thermal relaxation process is followed by the isothermal process from $C$ to $D$.     The total heat $Q_{\text{out}}$ dumped to the cold reservoir ($Q_\text{out}>0$ indicates heat flowing from system to cold reservoir) is hence contributed by two terms, with
\begin{equation}\label{Q_out}
  Q_{\text{out}}=T_c(S_C-S_D)+\langle E_{C^\prime} \rangle-\langle E_C \rangle,
\end{equation}
where $S_C$ and $S_D$ describe entropy of equilibrium states $C$ and $D$. In the same fashion,  the total heat absorbed from the hot reservoir, denoted by $Q_{\text{in}}$ ($Q_\text{in}>0$ indicates heat flowing from hot reservoir to system), is given by
\begin{equation}\label{Q_in}
  Q_{\text{in}}=T_h(S_B-S_A)+\langle E_A \rangle-\langle E_{A^\prime} \rangle.
\end{equation}
The efficiency of such a general quantum engine cycle is therefore $\eta_q=1-Q_\text{out}/Q_\text{in}$, {\it i.e.,}
\begin{equation}\label{efficiency}
    \eta_q =1-\frac{T_c(S_C-S_D)+\langle E_{C^\prime} \rangle-\langle E_C \rangle}{T_h(S_B-S_A)+\langle E_A \rangle-\langle E_{A^\prime}\rangle}.
\end{equation}
To compare the above efficiency $\eta_q$ with the Carnot efficiency $\eta_c\equiv 1-{T_c}/{T_h}$, we first
define $\Delta S_{B\rightarrow C}^{\text{total}}$ and $\Delta S_{D\rightarrow A}^{\text{total}}$, namely
the total entropy increase of the universe for the overall process $D\rightarrow A$ and $B\rightarrow C$.
It is straightforward to obtain
\begin{eqnarray}
\Delta S_{B\rightarrow C}^{\text{total}}&=&(S_C-S_B)- \frac{1}{T_c}[\langle E_C \rangle-\langle E_{C^\prime} \rangle]; \nonumber \\
\Delta S_{D\rightarrow A}^{\text{total}}&=& (S_A-S_D)-\frac{1}{T_h}[\langle E_A \rangle-\langle E_{A^\prime} \rangle].
\end{eqnarray}
By the second law of thermodynamics, both $\Delta S_{B\rightarrow C}^{\text{total}}$ and $\Delta S_{D\rightarrow A}^{\text{total}}$ cannot be negative.
Let us now rewrite Eq.~(\ref{efficiency}) as:
\begin{equation}\label{Efficiency1}
    \eta_q =1-\frac{T_c(S_B-S_D)+T_c \Delta S_{B\rightarrow C}^{\text{total}}}{T_h(S_B-S_D)-T_h\Delta S_{D\rightarrow A}^{\text{total}}}.
\end{equation}
Evidently then, if $\Delta S_{D\rightarrow A}^{\text{total}}$ and $\Delta S_{B\rightarrow C}^{\text{total}}$ in Eq.~(\ref{Efficiency1}) is zero, then $\eta_q$ would reduce exactly to the Carnot efficiency $\eta_c$.
In general, $\eta_q$ in Eq.~(\ref{Efficiency1}) is seen to be lower than $\eta_c$. In short, the second law of thermodynamics implies that,  the efficiency of a quantum heat engine cycle described above should be in general lower than, and can only reach in exceptional cases,  the Carnot efficiency.

{\it Constructing a quantum Carnot cycle} -- To construct a quantum Carnot cycle, one must specify the two unitary processes $B\rightarrow C^\prime$ and $D\rightarrow A^\prime$.  Reference \cite{Bender.00.JPAMG} first proposed to consider quantum adiabatic processes for this purpose, mainly based on reversibility considerations \cite{note}. Here we show that this intuitive construction is correct for a more fundamental reason related to the heat-to-work efficiency.

Before proceeding, we emphasize that adiabaticity in a quantum unitary process does not have the key feature of a thermal quasi-static adiabatic process in the Carnot cycle, i.e., the former does not result in equilibrium states in general but the latter does.  As a result, quantities such as temperature and thermodynamic  are usually ill-defined for states $C^\prime$ and $A^\prime$.  Consider then the expression of $\eta_q$ in Eq.~(\ref{efficiency}). With four equilibrium states $A$, $B$, $C$ and $D$ specified as in Fig.~\ref{Fig1},  only $\langle E_{C^\prime} \rangle$ and $\langle E_{A^\prime} \rangle$ may be varied by choosing different types of unitary processes $B\rightarrow C^\prime$ and $D\rightarrow A^\prime$.   For thermally isolated processes, there is no heat exchange and as such, we have
\begin{eqnarray}
    \langle E_{C^\prime} \rangle & = &\langle E_B \rangle+\langle W \rangle _{B \rightarrow C^\prime};\nonumber \\
    \langle E_{A^\prime} \rangle & = &\langle E_D \rangle+\langle W \rangle _{D \rightarrow, A^\prime}\label{miniwork}
\end{eqnarray}
where $\langle W \rangle _{B \rightarrow C^\prime}$ and $\langle W \rangle _{D \rightarrow A^\prime}$ represent the average work associated with $B\rightarrow C^\prime$ and $D\rightarrow A^\prime$. Remarkably, the minimal work principle \cite{Allahverdyan.05.PRE} then takes us to a definite choice. In particular, for a quantum state initially prepared as a Gibbs equilibrium distribution (this specific requirement can be loosened) and for fixed initial and final $\lambda$ values, a quantum adiabatic process (if implementable) is the one with the minimal average work.
So if $D\rightarrow A^\prime$ and $B\rightarrow C^\prime$ are indeed quantum adiabatic processes,
the minimal work principle ensures that the final mean energies $\langle E_{C^\prime} \rangle$ or $\langle E_{A^\prime} \rangle$ are minimized for fixed states $B$ and $D$.  Returning to the expression of $\eta_q$ in  Eq.~(\ref{efficiency}), minimized $\langle E_{C^\prime} \rangle$ and $\langle E_{A^\prime} \rangle$ then yield the highest possible efficiency $\eta_q$. It is for this efficiency consideration that the quantum analog of Carnot heat engines must consist of  two quantum adiabatic steps in addition to two isothermal steps.  To our knowledge, this is an important and previously unknown insight \cite{note}.

{\it Optimizing efficiency of quantum Carnot cycles} -- With quantum Carnot cycles constructed and justified as above, we next seek specific design principles to further optimize $\eta_q$.
The Hamiltonian of the working medium (when thermally isolated) is assumed to be $\hat{H}(\lambda)$ with energy levels $E_n(\lambda)$.  The values of $\lambda$ at $B$ and $D$, namely, $\lambda_B$ and $\lambda_D$,  are assumed to be given.  The focus question of this study is to show how to choose $\lambda$ at states $A$ and $C$, namely, $\lambda_A$ and $\lambda_C$, such that $\eta_q$ may be optimized.

It is interesting to first illustrate this optimization issue in systems possessing scale invariance \cite{low2,scale} with $\lambda$. In such an exceptional case,
\begin{equation}
\label{scale-invariance}
[E_{n}(\lambda_1)-E_{m}(\lambda_1)] = S(\lambda_1,\lambda_2)  [E_{n}(\lambda_2)-E_{m}(\lambda_2)].
\end{equation}
Examples of this situation include a harmonic oscillator, with $\lambda$ being as the harmonic frequency, a particle in a infinitely deep square-well potential \cite{Bender.00.JPAMG}, where $\lambda$ can be the width of the potential well, or simply a two-level quantum system \cite{prl2}. Consider then the adiabatic step from $B$ to $C^\prime$.  The initial state populations are given by
$P_B(n)=e^{-\beta_h E_n(\lambda_B)}/Z_{B}$ (throughout $Z$ represents equilibrium partition functions and $\beta$ represents the inverse temperature).  Upon reaching $C^\prime$, the final populations are still given by $P_{C^\prime}(n)=e^{-\beta_h E_n(\lambda_B)}/Z_{B}$ due to the assumed quantum adiabaticity. Now given the assumed scale-invariance in Eq.~(\ref{scale-invariance}),
one can always define an effective temperature $T_{\text{eff}}$ to reinterpret $P_{C^\prime}(n)$, namely,
\begin{equation}
P_{C^\prime}(n)=e^{-\beta_h E_n(\lambda_B)}/Z_{B}=e^{-\beta_{\text{eff}} E_n(\lambda_C)}/Z_{C},
\end{equation}
where $\beta_{\text{eff}}\equiv 1/(k_B T_{\text{eff}})=S(\lambda_B, \lambda_C) \beta_h$.  That is, state $C^\prime$ has no difference from an equilibrium state with temperature $T_{\text{eff}}$ and Hamiltonian $\hat{H}(\lambda_C)$. If we now choose $\lambda_C$ to guarantee that $T_{\text{eff}}=T_c$, then state $C^\prime$ is already in thermal equilibrium with the cold reservoir at $T_c$.  The relaxation process from $C^\prime$ to $C$ as illustrated in Fig.~\ref{Fig1} is no longer needed, resulting in $S_C=S_B$, $\langle E_C\rangle = \langle E_{C^\prime}\rangle$ and hence $\Delta S_{B\rightarrow C}^{\text{total}}=0$.  Exactly the same analysis applies to the adiabatic process $D\rightarrow A^\prime$.
That is, by choosing an appropriate value of $\lambda_A$,  we can set
$\Delta S_{D\rightarrow A}^{\text{total}}=0$.  According to the expression of $\eta_q$ in Eq.~(\ref{Efficiency1}),  $\eta_q$ then yields the standard Carnot efficiency. This result also offers a clear perspective to explain why the Carnot efficiency can be obtained in some early studies of quantum heat engines ~\cite{Bender.00.JPAMG,Arnaud.08.PRE}.

We next lift the above scale-invariance assumption and proceed with optimizing $\eta_q$, by optimizing $Q_{\text{out}}$ and $Q_{\text{in}}$.
We first rewrite $Q_{\text{out}}$ and $Q_{\text{in}}$ in Eqs.~(\ref{Q_out}) and (\ref{Q_in}) as the following:
\begin{eqnarray}\label{Q2}
    Q_{\text{out}} & =&\langle E_{C^\prime} \rangle-\langle E_D \rangle-k_B T_c\ln \frac {Z_D}{Z_C}, \nonumber \\
    Q_{\text{in}}  & =& \langle E_B \rangle-\langle E_{A^\prime} \rangle+k_B T_h\ln \frac {Z_B}{Z_A}.
\end{eqnarray}
Interestingly, with states $B$ and $D$ fixed, only $\lambda_C$ may affect $Q_{\text{out}}$
via $\langle E_{C^\prime} \rangle$ and $Z_C$; while only $\lambda_A$ may affect $Q_{\text{in}}$ via $\langle E_{A^\prime} \rangle$ and $Z_A$. That is, to optimize $\eta_q$, minimizing $Q_{\text{out}}$ and maximizing $Q_{\text{in}}$ can be executed separately, which is a considerable reduction of our optimization task.  For this reason, below we focus on minimizing
$Q_{\text{out}}$ and the parallel result concerning $Q_{\text{in}}$ directly follows.

Accounting for quantum adiabaticity that maintains populations on each quantum level, one has
\begin{equation}\label{E^prime_C}
  \langle E_{C^\prime} \rangle=\frac{1}{Z_B}\sum_n e^{-\beta_h E_n(\lambda_B)}E_n(\lambda_C).
\end{equation}
Note again that the level populations $\frac{1}{Z_B} e^{-\beta_h E_n(\lambda_B)}$ used above are in general not an equilibrium Gibbs distribution associated with $\hat{H}(\lambda_C)$.  Using Eqs.~(\ref{Q2}) and (\ref{E^prime_C}), we arrive at
\begin{equation}\label{dQ_out}
\begin{split}
\frac{\partial Q_{\text{out}}}{\partial \lambda_C} & = \frac{\partial \langle E_{C^\prime} \rangle}{\partial \lambda_C}+k_B T_c\frac{\partial (\ln Z_C)}{\partial \lambda_C}\\
                 & =\sum_n \left[\frac{e^{-\beta_h E_n(\lambda_B)}}{Z_B}-\frac{e^{-\beta_c E_n(\lambda_C)}}{Z_C}\right] \frac{\partial E_n(\lambda_C)}{\partial \lambda_C}.
\end{split}
\end{equation}
The minimization of $Q_{\text{out}}$ requires the condition ${\partial Q_{\text{out}}}/{\partial \lambda_C}=0$, which indicates that
\begin{equation}\label{dQ_out1}
  \sum_n \left[\frac{e^{-\beta_h E_n(\lambda_B)}}{Z_B}-\frac{e^{-\beta_c E_n(\lambda_C)}}{Z_C}\right] \frac{\partial E_n(\lambda_C)}{\partial \lambda_C}=0.
\end{equation}
Viewing the linear response in energy to a variation in $\lambda$ as a generalized force, we define a general force operator $ \hat{\cal F}_\lambda\equiv -\frac{\partial \hat{H}(\lambda)}{\partial \lambda}$. Then the condition in Eq.~(\ref{dQ_out1}) can be cast in the following compact form,
\begin{equation}\label{dQ_out2}
  \langle \hat{\cal F}_{\lambda_C}\rangle_C= \langle \hat{U}^{\dagger}_{B\rightarrow C} \hat{\cal F}_{\lambda_C}\hat{U}_{B\rightarrow C} \rangle_B,
\end{equation}
where $\hat{U}_{B\rightarrow C}$ is the unitary transformation that transforms an arbitrary $n$th eigenstate of $\hat{H}(\lambda_B)$ to the $n$th eigenstate of $\hat{H}(\lambda_C)$.
That is, the expectation value of a generalized force operator at $\lambda_C$ over equilibrium state $C$ should be identical with that of a mapped force operator over equilibrium state $B$.
Needless to say, the condition for $Q_{\text{in}}$ to be maximized is given by
\begin{equation}\label{dQ_in1}
  \langle \hat{\cal F}_{\lambda_A} \rangle_A=\langle \hat{U}^{\dagger}_{D\rightarrow A} \hat{\cal F}_{\lambda_A}  \hat{U}_{D\rightarrow A} \rangle_D,
\end{equation}
where $\hat{U}_{D\rightarrow A}$ transforms an arbitrary $n$th eigenstate of $\hat{H}(\lambda_D)$ to the $n$th eigenstate of $\hat{H}(\lambda_A)$.

Unlike a previous interesting suggestion \cite{Quan.14.PRE}, the two explicit conditions in Eq.~(\ref{dQ_out2}) and (\ref{dQ_in1}) to optimize $\eta_q$ are not about matching the mean energy between states $C^\prime$ and $C$ ($A^\prime$ and $A$). Attempts to match information entropy between states $C^\prime$ and $C$ ($A^\prime$ and $A$) do not optimize $\eta_q$, either.  Rather, the conditions found here are about a
more subtle and more involving
matching of the expectation values of some generalized force operators through equilibrium states.  We now take Eq.~(\ref{dQ_out1}) as the example to digest the optimization conditions.
For the exceptional case of a scale-invariant medium,  due to the existence of a $\beta_{\text{eff}}=\beta_c$ at $C^\prime$, one can achieve $\frac{e^{-\beta_h E_n(\lambda_B)}}{Z_B}-\frac{e^{-\beta_c E_n(\lambda_C)}}{Z_C}=0$ for arbitrary $n$.  Then
 Eq.~(\ref{dQ_out1}) can be easily satisfied, independent of the details of $\frac{\partial E_n(\lambda_C)}{\partial \lambda_C}$.  For a general working medium, the condition of Eq.~(\ref{dQ_out1}) may be still satisfied after setting the sum of all the terms $\frac{\partial E_n(\lambda_C)}{\partial \lambda_C} \left[\frac{e^{-\beta_h E_n(\lambda_B)}}{Z_B}-\frac{e^{-\beta_c E_n(\lambda_C)}}{Z_C}\right]$ to zero.

{\it Numerical examples} -- We adopt a simple model system that is not scale-variant with $\lambda$, with
 $E_n(\lambda)=\lambda n +\alpha n^2 + {\rm const}$ (all variables in dimensionless units). For our purpose here there is no need to specify the explicit form of the Hamiltonian.  If $\alpha$ is comparable to $\lambda$, then the ratio  ${[E_n(\lambda_1)-E_m(\lambda_1)]}/[E_n(\lambda_2)-E_m(\lambda_2)]$ does depend strongly on $n$ and $m$, a clear sign of breaking the scale invariance.  Cases of a very small $\alpha$ would resemble the behavior of a harmonic oscillator at low temperatures.  To guarantee quantum adiabaticity,   we exclude cases with level crossings.
This is achieved by requiring $\lambda > -\alpha$ such that $E_n>E_m$ if $n>m$.  Other physical considerations for the cycle to operate as a meaningful heat engine suggests that $\lambda_A$ should be the largest and $\lambda_C$  the smallest among $\lambda_A$, $\lambda_B$, $\lambda_C$, and $\lambda_D$.
Our computational details confirm that minimization of $Q_{\text{out}}$ and maximization of $Q_{\text{in}}$ indeed occur precisely at those locations predicted by Eq.~(\ref{dQ_out2}) and (\ref{dQ_in1}).  Optimization of $\eta_q$ as we outlined above theoretically is hence indeed doable.

\begin{figure}[htp]
\centering
\subfigure[]{
\label{Fig2a}
\includegraphics[width=7.4cm]{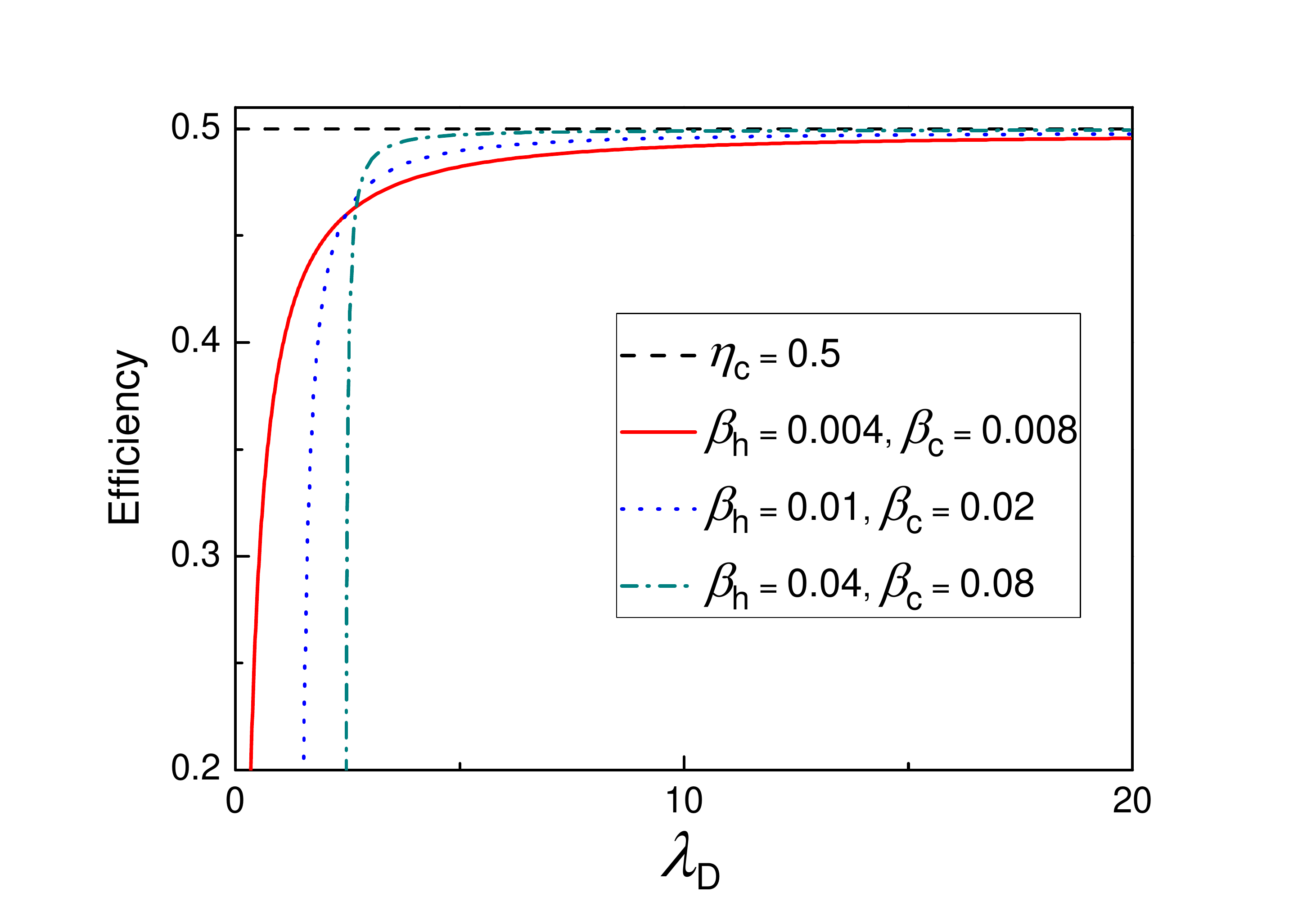}}
\subfigure[]{
\label{Fig2b}
\includegraphics[width=7.4cm]{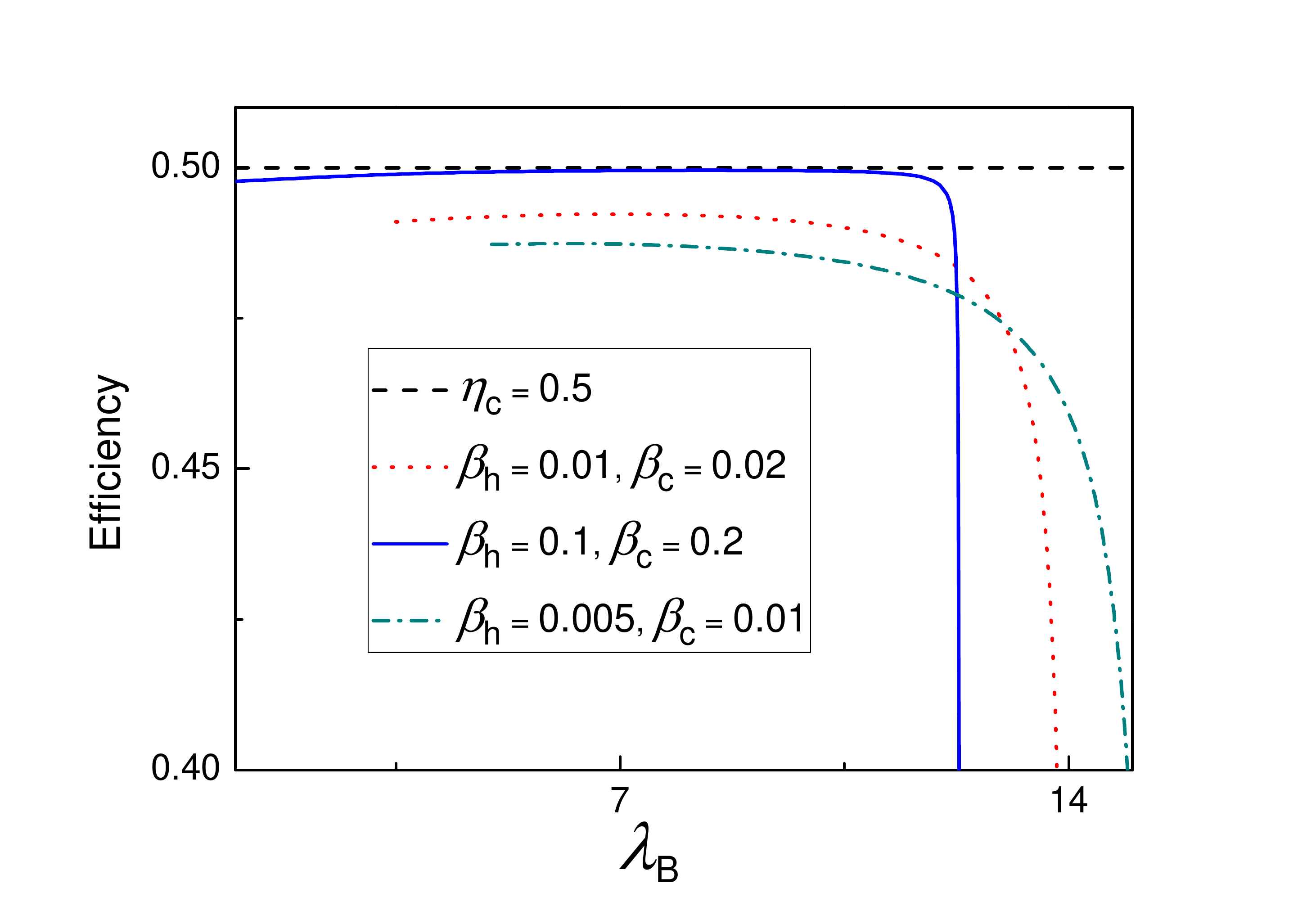}}
  \caption{(Color online) Behavior of  heat-to-work efficiency $\eta_q$ optimized under given
   $\lambda_B$, $\lambda_D$, $T_c$ and $T_h$, with $\lambda_A$ and $\lambda_C$ chosen according to
   Eq.~(\ref{dQ_out2}) and Eq.~(\ref{dQ_in1}).
   (a) $\eta_q$ vs $\lambda_D$ if $\lambda_B=6.0$ for three sets of $T_c$ and $T_h$.
    and (b) $\eta_q$ vs $\lambda_B$ if $\lambda_D=6.0$, also for three sets of $T_c$ and $T_h$. In all the shown cases the Carnot efficiency $\eta_c=0.5$ (top dashed curve).   All variables are in dimensionless units, and the energy levels of the working medium
   are assumed to be $E_n(\lambda)=\lambda n +\alpha n^2$ with $\alpha=0.1$.}
\label{Fig2}
\end{figure}

The rather specific conditions to optimize $\eta_q$ (under fixed $\lambda_B$ and $\lambda_D$)
indicates that the  $\eta_q$  thus optimized will be highly system specific. To see this, we present in Fig.~\ref{Fig2} optimized $\eta_q$ as a function of $\lambda_D$ ($\lambda_B)$ with fixed $\lambda_B$ ($\lambda_D$), under different temperatures $T_c$ and $T_h$.
 From Fig.~\ref{Fig2a} it is seen that the optimized $\eta_q$ can be way below, but nevertheless quickly approaches, the Carnot efficiency $\eta_c$ as $\lambda_D$ increases.  This is because a larger $\lambda_D$ leads to an even larger $\lambda_A$ due to the optimization requirement, both facts pushing the system closer to a scale-invariant system under fixed temperatures.  Note also that even though the three $\eta_q$ curves in Fig.~\ref{Fig2a} are for the same ratio $T_c/T_h$, their $\eta_q$ values are much different. This shows that the optimized $\eta_q$ is no longer a simple function of $T_c/T_h$, but a function of both $T_c$ and $T_h$.
Figure ~\ref{Fig2b} shows that our optimized $\eta_q$ may not be always a monotonous function of $\lambda_B$ with a fixed $\lambda_D$. Interesting effects of $T_c$ and $T_h$ under a common $T_c/T_h$ are again observed there.
The shown ranges of $\lambda_B$ or $\lambda_D$ in Fig.~\ref{Fig2}   vary with the chosen temperatures because we exclude level crossing situations. The sharp change of $\eta_q$ in Fig.~\ref{Fig2a} [Fig.\ref{Fig2b}] with a decreasing (increasing) of $\lambda_D$ ($\lambda_B$) under a given $\lambda_B$ ($\lambda_D$) is simply because
the optimized cycle is about to cease to operate as a heat engine (which requires $Q_{\text{in}}>Q_{\text{out}}>0$).

{\it Discussions and conclusions} -- Several previous studies investigated quantum Otto cycles \cite{Rezek.06.NJP,Abah.12.PRL,Dario,gongpre13}  consisting of two thermally isolated steps and two isochoric processes that are simply relaxation processes with a hot or a cold reservoir. It is clear that such quantum Otto cycles can be regarded as a special case of the quantum heat engine cycles considered here, without the isothermal process $A\rightarrow B$ or $C\rightarrow D$. That is, by setting $\lambda_{A}=\lambda_B$ and $\lambda_{C}=\lambda_D$ in Fig.~\ref{Fig1}, we obtain the quantum Otto cycles.  One can now also justify the use of quantum adiabatic steps to construct energy efficient quantum Otto cycles, using the  minimum work principle again \cite{Allahverdyan.05.PRE}. But more importantly, because in our efficiency optimization under fixed $\lambda_B$ and $\lambda_D$, the obtained $\lambda_{A}$ ($\lambda_{C})$  in general differs from $\lambda_B$ ($\lambda_D$), one deduces that the optimized $\eta_q$ here is in general higher than the efficiency of the corresponding quantum Otto cycles.  This fact strengthens the importance of quantum Carnot cycles we have justified and optimized.

Our analysis of the two adiabatic steps does not really demand the two steps to be executed slowly. That is, so long as the final-state populations are consistent with those expected from the quantum adiabatic theorem, then all our results will be valid.  This understanding encourages the use of shortcuts to adiabaticity \cite{scale,shortcut,gongpre13,campo13,Tu} or even an optimal control approach \cite{Xiao} to implement the quantum Carnot cycles within a shorter time scale, thus boosting the heat engine power.

In conclusion,  using minimal assumptions about the concepts of work and heat in few-body systems, we have shown how to construct and optimize the quantum analog of Carnot cycles at the nanoscale.  The heat-to-work efficiency can be optimized if two conditions regarding some generalized force operators evaluated at some equilibrium states are met.  In general the optimized efficiency is system specific, lower than the Carnot efficiency, and dependent upon both temperatures of the cold and hot reservoirs.


\begin{thebibliography}{99}
\bibitem{note} The reversibility arguement in Ref.~\cite{Bender.00.JPAMG} to support the use of
quantum adiabatic processes is not sufficient because
 an arbitray unitary process in an isolated few-body quantum system can be easily reversed. Similarly, Ref.~\cite{prl2} explained the use of quantum adiabatic processes by the principle of maximum work (which is also based on the notion of reversibility in thermodynamics processes), implicitly and again intuitively assuming that quantum adiabatic processes should replace quasi-static adiabatic steps in thermodynamics without a formal justification.
\end{thebibliography}
\end{document}